\documentclass[aps,pra,twocolumn,twoside,a4]{revtex4}
\usepackage{hyperref}
\usepackage{bbm}
\usepackage{amsmath, amssymb}
\usepackage{color}
\usepackage{graphicx,epsfig}
\usepackage{pifont} 
\usepackage{amssymb}

\newcommand{\tr}{{\rm tr}}

\newcommand{\de}{\delta}
\newtheorem{defi}{Definition}\def\DE{\begin{defi}}\def\ED{\end{defi}}
\newtheorem{lemma}{Lemma}\def\LE{\begin{lemma}}\def\EL{\end{lemma}}
\newtheorem{theo}{Theorem}\def\LE{\begin{theo}}\def\EL{\end{theo}}

\newcommand{\figsize}{0.4}

\begin{document}

\title{Thermodynamics of discrete quantum processes}

\author{Janet Anders}
\email{janet@qipc.org}
\affiliation{Department of Physics and Astronomy, University College London, Gower Street, London WC1E 6BT, United Kingdom.}

\author{Vittorio Giovannetti}
\email{v.giovannetti@sns.it}
\affiliation{NEST, Scuola Normale Superiore, and Istituto Nanoscienze - CNR,    \\
 Piazza dei Cavalieri 7, I-56127 Pisa, Italy.}


\begin{abstract}

We define thermodynamic configurations and identify two primitives of discrete quantum processes between configurations for which heat and work can be defined in a natural way. This allows us to uncover a general second law for any discrete trajectory that consists of a sequence of these primitives, linking both equilibrium and non-equilibrium configurations. Moreover, in the limit of a discrete trajectory that passes through an infinite number of configurations, i.e. in the reversible limit, we recover the saturation of the second law. Finally, we show that for a discrete Carnot cycle operating between four configurations one recovers Carnot's thermal efficiency.

\end{abstract}



\maketitle

\section{Introduction}

The intuitive meaning of \emph{heat} and \emph{work} in thermodynamics is that of two types of energetic resources, one fully controllable and useful, the other uncontrolled and wasteful. An impressive effort has been devoted to provide a consistent mathematical characterisation of these notions within a quantum mechanical description of physics \cite{Mukamel03,Talkner2007,ESPOSITORMP,Campisi2009,Esposito2010,CAMPISI2011,Horodecki2011}. This is a challenge since in contrast to other thermodynamic quantities, such as internal energy and entropy, {heat} and {work} are not properties of  individual states of a system. They are defined for \emph{continuous processes} connecting different states~\cite{KURCHAN,TASAKI,Talkner2007}, implying that their statistical fluctuations cannot be described in terms of  a single system observable. Two-point correlation functions characterising the correlations along process paths are required - a problematic territory for quantum mechanics where definite trajectories cannot be fixed unless the system is continuously measured. Resolving these issues has been the topic of a number of publications that have formulated quantum trajectory approaches  \cite{ESPOSITORMP,CAMPISI2011,KURCHAN,TASAKI,Talkner2007,TALKNER2009,Horowitz2012}.

In contrast here we focus on the \emph{mean} values of  heat and work where the analysis simplifies but still requires careful thought. We will adopt the identification of the system's internal energy with $U(\rho) = \tr[\rho \, H]$ where $\rho$ is the density matrix describing the state of the system at given time, and $H$ is its instantaneous Hamiltonian. Clearly, a proper definition of this Hamiltonian is in general problematic! If the system is coupled to an environment the non-equilibrium behaviour of a general open system makes the definition of the system's Hamiltonian ambiguous \cite{TALKNER2009,ALLA,WEIM,HORH,Deffner2008} both, mathematically and experimentally. 
Ultimately the choice of the Hamiltonian one assigns to the system must rely on the set of operations and observables one can access experimentally.
In many situations of physical interest $H$ can be identified with the bare system Hamiltonian or an effective system Hamiltonian that incorporates the effect of the environment. 

While the environment degrees are in principle uncontrolled, full control can be exerted over the temporal ``variation" of the system Hamiltonian. For instance, the size of a container in which steam is pumped can be freely chosen and a piston can be attached to the container that can push the wheels of a train. A formal definition of mean heat and work is then obtained by considering an infinitesimal change of the internal energy  
\begin{eqnarray} \label{firsteq}
	{d}U = {d}\tr[ \rho \,  H ] =  \tr[\rho \, {d}H] + \tr[ {d}\rho \, H]\;, 
\end{eqnarray} 
associated with the time evolution of the system which brings its density matrix from $\rho$ to $\rho + d\rho$ while the  Hamiltonian varies from $H$ to  $H + d H$. The origin for the change of $\rho$ may here be due to both, the variation of $H$ induced by the experimenter and by the dynamics due to the coupling with the environment. 
The possibility of externally controlling $H$ suggests to identify the first term on the r.h.s. of Eq.~(\ref{firsteq}) with the average work done by the experimenter during the evolution. The second term describes the internal energy change due to a reconfiguration of the system, i.e. a variation of the system's density matrix. This is an energy contribution over which the experimenter has no direct control and this is why it is associated with heat. The infinitesimal average \emph{heat absorbed by the system} and the infinitesimal average \emph{work done on the system} \cite{PW,LENARD,ALICKI,KR,TASAKI2,KIEW,ALLA,ESPOSITO,HENRICH} are therefore defined as
\begin{eqnarray} \label{eq:heatwork}
	{\de}Q := \tr[ {d}\rho \, H] \quad \mbox{and} \quad {\de}W := \tr[ \rho \,  {d}H]\;,
\end{eqnarray}
with the symbol $\de$ indicating that heat and work are in general no full differentials, i.e. they do not correspond to observables.

While the first law of thermodynamics states that the sum of the two average energy types is the average internal energy,
\begin{eqnarray} \label{eq:firstlaw}
	{d}U = {\de}Q + {\de}W\;,
\end{eqnarray}
the \emph{split} into these two types of energies is crucial as it allows the formulation of the second law of thermodynamics. A fundamental law of physics, it sets limits on the work extraction of heat engines and establishes the notion of \emph{irreversibility} in physics. The second law can be phrased in form of Clausius' inequality:
\begin{eqnarray}
	T d S \ge  \de Q, 
\end{eqnarray}
stating that the change in a system's entropy must be larger than the average heat absorbed by the system during a process. While the first law of thermodynamics is fundamental for any process, the second law was originally stated for processes that \emph{start and end in equilibrium}. Recently, the non-equilibrium work relation due to Jarzynski has been used to argue that the second law should also hold for any process starting from equilibrium, at temperature $T$, but ending in an arbitrary non-equilibrium state \cite{JAR}. However, no conclusive argument has yet established the most general set of dynamical processes that obey the Clausius inequality \cite{Esposito2011}.  

Extending the infinitesimal scenario to \emph{finite, continuous processes} in which the temporal evolution of $\rho(t)$ and $H(t)$ in time $t$ is known, the mean heat and work can be found by integrating over the trajectory taken from $\rho(0)$ and $H (0)$ to $\rho(\tau)$ and $H(\tau)$, i.e. 
\begin{eqnarray} 
	Q  &: =& \int_{0}^{\tau} \, {d}t \, \tr[ \dot \rho(t) \,  H(t)] \;,  \label{eq:Q} \\ 
	W &: =& \int_{0}^{\tau} \, {d}t \, \tr[ \rho(t) \,  \dot {H}(t)]\;, \label{eq:Q+W}
\end{eqnarray}
while the first law becomes
\begin{eqnarray} \label{eq:1stlaw}
	\Delta U : =  \int_{0}^{\tau} \, {d}t \, {{d}\over {d}t} \tr[ \rho(t) \,  H(t)] =Q + W\;.
\end{eqnarray}
The mathematical consistency of the above expressions and their compatibility with the  predictions of thermodynamics have been verified for many models, for example, for processes that are induced by Markovian master equations~\cite{ALICKI}. 

There are two paradigmatic examples of \emph{all work} and \emph{all heat} processes that we introduce here and which will become important in the later part of the manuscript. The first process is a unitary process, which we will also refer to as \emph{closed}, where the (non-equilibrium) evolution of the state is given by the Schr\"odinger equation, 
\begin{eqnarray} \label{eq:unitary}
	\dot \rho (t) = - {i\over \hbar} [H(t), \rho(t)]\;. \label{inUNI}
\end{eqnarray}
Mean heat and work are then
\begin{eqnarray} \nonumber 
	 Q_{unitary} &=& - {i\over \hbar} \int_{0}^{\tau} \, {d}t \, \tr[ [H(t), \rho(t)] \, H(t)]  = 0\;, \\
 	W_{unitary} &=& \Delta U\;, \label{example1}
\end{eqnarray}
consistent with the physical intuition that \emph{no} heat has been provided to the system during  the evolution.
The second example is a system that evolves through the action of a dissipative, i.e. \emph{open}, Markov process via a master equation~\cite{PETRU},
\begin{eqnarray}
	\dot\rho(t) = -\frac{i}{\hbar}  [H,\rho(t)] + {\cal L}(\rho(t))\;, \label{ddf}
\end{eqnarray}
with ${\cal L}$ being the  dissipative Lindblad term. Under the assumption that the typical time scales associated with the time-independent $H$ are much shorter than those associated with ${\cal L}$ we can treat the system as almost isolated and use Eq.(\ref{firsteq}) to compute its internal energy. In this limit Eq.~(\ref{eq:Q+W}) is valid with the Hamiltonian just being the time-independent $H$, 
\begin{eqnarray}
	W_{dissipative}  &=& 0, \nonumber \\
	Q_{dissipative} &=& \tr [\rho(\tau) \, H ] - \tr [\rho(0) \, H] = \Delta U, \,\, \label{IDENTITY111}
\end{eqnarray}
which is in full agreement with the physical intuition that no work has been performed on the system. 

While these examples constitute special cases of continuous processes the heat and work in a general process depend intimately on the exact details of the process. However, the caveat with this viewpoint is that in most real life applications one does \emph{not} know what the dynamics of the state of the system is nor what the appropriate local Hamiltonian is at all times. Importantly, this is not just due to our ignorance of what happens at the quantum level. Quantum physics  has strong fundamental limitations on what we \emph{can know}  without choosing a measurement apparatus, measuring the system and interpreting the data. Moreover, if the system is indeed measured then the experimenter's choice of what degrees of freedom she actually measures will effect what the \emph{measured} heat and work will be. In other words, we propose that there is no \emph{one} average heat and work for a particular process, there are different sensible outcomes to this question and the answer depends on the choice of system Hamiltonians in time, $H(t)$, that corresponds to specific measurement choices. 

The aim of this paper is to show that it is possible to formulate a general second law independently of these choices. To achieve this we will depart entirely from the traditional continuous trajectory approach and propose a rather drastic but pragmatic change of perspective. We develop a consistent framework of mean heat and work for  \emph{discrete thermodynamic processes}. The rationale for this approach is that while the true process is continuous, observations we make on the system are almost always discrete. (We will neglect here the possibility of monitoring through continuous weak measurements.)  For discrete snapshots of the dynamics, we find that by decomposing the transition into possible sequences of two fundamental primitives, it is possible to define heat and work for the discrete process in a way that is experimentally and mathematically clear. This allows us to establish a general second law for discrete processes between equilibrium and non-equilibrium states and the analysis of a discrete Carnot cycle, where we uncover the usual Carnot efficiency. 

The paper is structured as follows. In Sec.~\ref{sec:entropy} we review the traditional perspective on the second law and the definition of entropy. In Sec.~\ref{sec:configuration} we define the dynamical configuration space of a system that allows us to formulate a notion of two primitives for discrete processes in Sec.~\ref{sec:DUTs+DTTs}, the discrete unitary and discrete thermalising transformations (DUTs and DTTs). Sec.~\ref{sec:2ndlaw} contains the main results of the paper. First we show that entropic inequalities when applied to discrete trajectories formed by concatenating DUTs and DTTs yields the second law of thermodynamics in the Clausius formulation. We then derive two consequences: We find the minimum and maximum heat for a single DUT and DTT sequence and prove the existence of a discrete trajectory, formed by sequences of DUTs and DTTs, that connects two given thermodynamical configurations while asymptotically saturating the Clausius inequality. Finally  we identify a discrete trajectory that connects the same initial and final configurations as the continuous trajectory through a sequence of DUTs and DTTs, and which approximates the continuous heat. In Sec.~\ref{sec:efficiency} we derive the thermal efficiency of a discrete cycle, the Carnot efficiency,  and conclude in Sec.~\ref{sec:Conc}. 

\section{Entropy and the second law}\label{sec:entropy}

In 1865 Clausius established that the overall heat flow in any cyclic, reversible process vanishes, implying that the integral over any non-cyclic process must be path independent. This led him to define the state function \emph{entropy}, $S$, and the entropy change, $\Delta S$, between the final and initial point of a reversible process,
\begin{equation} \label{eq:reversible}
	\oint_{rev} {\de Q \over T} = 0 \quad \Rightarrow \quad	\int_{rev} {\de Q \over T} = : \Delta S.
\end{equation}
Clausius also showed that any cyclic process, reversible or irreversible, obeys 
\begin{equation}
	\oint {\de Q \over T} \le 0.
\end{equation}
This relation is the basis for a formulation of the second law of thermodynamics, known as the \emph{Clausius-inequality}. It is a statement for all thermodynamic processes, not just cyclic ones, that start from equilibrium at temperature $T$,
\begin{equation} \label{eq:Clausius}
	\int {\de Q \over T} \le \Delta S, 
\end{equation}
and it simplifies to $Q \le T \Delta S$ when the system interacts with a bath at constant temperature, $T$.  In this form Clausius' inequality establishes the existence of an upper bound to the heat received by the system.

Clearly, Clausius's goal was to characterise different forms of \emph{energy} and their interconversion. However, by formulating the second law of thermodynamics he defined a new quantity: entropy. In contrast, in modern information theory the focus is on the \emph{state} of a system. Entropy is here used as the central physical quantity to measure the amount of information of a state, while heat and work, and energy in general, have no well-defined purpose for the interpretation of information processing. This opens the possibility of turning Clausius' original argument around! It allows us to use the entropy change in discrete quantum processes to \emph{define} the average heat, and work. Before we proceed, let us first highlight that non-trivial entropy bounds exists for any process between two states.

A state $\rho$ describes an amount of information, quantified by the \emph{von Neumann entropy}, $S (\rho) = - \tr[\rho \, \ln \rho]$. The evolution of a quantum system from an initial state  $\rho_i$  to a final state $\rho_f$ through an arbitrary process, or \emph{quantum channel}, has a meaningful associated entropy change,
\begin{eqnarray} \label{eq:entropychange}
	\Delta S (\rho_i, \rho_f)&=& -\tr[\rho_f \ln \rho_f] + \tr[\rho_i \ln \rho_i] ,
\end{eqnarray}  
which quantifies the change of the encoded amount of information. The entropy change is non-trivially bounded from above and below by virtue of the positivity of the \emph{relative entropy} (classically Kullback-Leibler divergence \cite{KL}). The relative entropy, $S(\rho_1||\rho_2)$, between two states $\rho_1$ and $\rho_2$ characterises the number of additional bits required to encode $\rho_1$ when using the diagonal basis of  $\rho_2$, rather than the diagonal basis of $\rho_1$. It is defined as~\cite{PETZ}
\begin{eqnarray} \label{defrelative}
  	S(\rho_1 \| \rho_2) := \tr [ \rho_1 \ln \rho_1] - \tr [\rho_1 \ln \rho_2]\;,
\end{eqnarray} 
and is a positive quantity
\begin{eqnarray} 
  	S(\rho_1 \|  \rho_2) \ge 0 \quad \mbox{with equality iff} \quad \rho_1 = \rho_2.
\end{eqnarray} 
Intuitively, the relative entropy is similar to a distance measure, however, it is important to keep in mind that it is asymmetric  $S(\rho_1 \| \rho_2) \not = S(\rho_2 \| \rho_1)$. Rewriting the entropy change, Eq.~(\ref{eq:entropychange}), in two ways
\begin{eqnarray}
	\Delta S (\rho_i, \rho_f) &=& -  \tr [ \Delta \rho \, \ln \rho_f ]  +  S(\rho_i\| \rho_f) \\
		 &=&  -  \tr [ \Delta \rho \, \ln \rho_i]  - S(\rho_f\| \rho_i),   \label{eq:V-entropy}
\end{eqnarray}  
a lower and upper bound on the entropy change emerge
\begin{eqnarray} \label{eq:entropybounds}
	-  \tr [ \Delta \rho \, \ln \rho_i] \geqslant \Delta S(\rho_i,\rho_f) \geqslant  -  \tr [ \Delta \rho \, \ln \rho_f] .
\end{eqnarray}  
From the information theory point of view, bounds on the entropy change are important in their own right as they characterise how much information is lost or gained. 

If we now assume the special case that $\rho_f$ is a thermal state for the Hamiltonian $H_f$ at an inverse temperature $\beta_f$ then the lower bound becomes
\begin{equation} \label{eq:entropy}
 	\Delta S (\rho_i, \rho_f) \geqslant  \beta_f  \, \tr [ \Delta \rho \, H_f].
\end{equation}  
Interpreting $\tr [ \Delta \rho \, H_f]$ as the heat of the discrete process, the above expression would constitute exactly the second law of thermodynamics. This is exactly what we will pursue in Sec.~\ref{sec:DUTs+DTTs}, e.g. in Eq. (\ref{DTTheat}). 

Interestingly, from (\ref{eq:entropybounds}) it is apparent that also an upper bound on the entropy change exists that is rarely discussed in the literature. This maximum value of the entropy change is enforced to ensure that any reverse process, from $\rho_f$ to $\rho_i$, also obeys the second law \footnote{Indeed the upper bound can be casted in the form of the lower bound by  simply reversing the role of the input and of the output configurations: i.e. $-  \tr [ \Delta \rho \, \ln \rho_i] \geqslant \Delta S(\rho_i,\rho_f) $
becomes $ \Delta S(\rho_f,\rho_i) \geqslant  - \tr [ \Delta^{(R)}\rho \, \ln \rho_f],$ with $\Delta^{(R)}\rho=\rho_i-\rho_f=- \Delta \rho$ being the state increment for the reverse process.}.

\section{Dynamical configuration space} \label{sec:configuration} 

\begin{figure}[t]
	\begin{center}
	{\includegraphics[width=\figsize \textwidth]{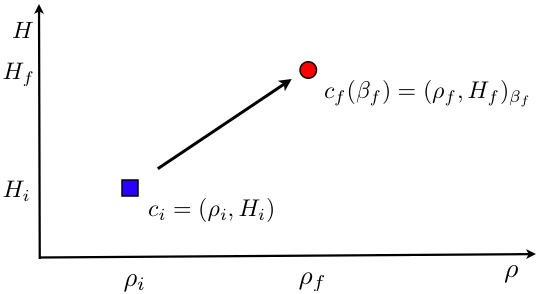}}
\caption{As a visual aid, points in dynamical configuration space are depicted in a $(\rho, H)$-coordinate system. A non-thermal configuration $c_i = (\rho_i, H_i)$ is shown as a blue square and a thermal configuration $c_f (\beta_f) = (\rho_f, H_f)_{\beta_f}$ is shown as a red circle. The discrete trajectory from $c_i$ to $c_f (\beta_f)$ is indicated by the black arrow. }
\label{fig:config}
\end{center}
\end{figure}

To assist our discussion of discrete quantum processes we introduce the concept of configuration space, following the spirit of~\cite{PW,LENARD,JANZ}, and propose a graphical representation for that space, see Fig.~\ref{fig:config}.
\begin{defi}
	Let $S$ be the quantum system under investigation, with ${\cal H}_S$ its Hilbert space, ${\cal L}({\cal H}_S)$ the set of linear operators on ${\cal H}_S$,  and $\mathfrak{S} ({\cal H}_S)\subset {\cal L}({\cal H}_S)$ the set of density matrices on~${\cal H}_S$. We define the \emph{dynamical configuration space} ${\cal C}({\cal H}_S)$ of $S$ as the set  formed by the pairs $(\rho, H) = c$ with $\rho \in \mathfrak{S} ({\cal H}_S)$ a density matrix 
\footnote{Density matrices are hermitian and positive operators with normalised spectrum, $\tr[\rho] =1$.}
and $H\in  {\cal L}({\cal H}_S)$ a Hermitian operator on ${\cal H}_S$ whose spectrum is bounded from below. Points in the dynamical configuration space $c$ are called ``configurations'' to distinguish them from ``states'', $\rho$.
\end{defi}
The evolution of the system is described by discrete  trajectories in ${\cal C}({\cal H}_S)$:
\begin{defi}
	A \emph{discrete trajectory}  ${\cal T}$ is defined as an ordered list of elements  of  ${\cal C}({\cal H}_S)$ that describes the succession of configurations, with each element $(\rho, H)$ containing {both} the density matrix $\rho$ of $S$ {and} the local Hamiltonian  $H$ of $S$ at that specific instance of the evolution. 
\end{defi}

We stress that both $\rho$ and $H$ of a configuration point $c\in {\cal C}({\cal H}_S)$ have a clear experimental meaning. $\rho$ is the density matrix that one would reconstruct by state tomography, i.e. the preparation of many copies of the same state $\rho$ and the full tomographic measurement of its properties. 
The Hamiltonian of the system, $H$, is determined by the set of projective measurements $\{ M_j\}_j$ the experimenter performs on the system to ``measure the energy'' together with the interpretation of the corresponding energy eigenvalues, $E_j$, so that $H= \sum_j E_j \, M_j$. (The choice of the measurement and interpretation can be motivated by a process tomography on the Hamiltonian at any point in time. For this the system needs to be decoupled from the rest of the universe at that instance and evolve for a complete set of states for a short time interval $\tau$ through the action of $H$. By measuring the final states of the evolution the unitary $e^{-i H \tau}$ and hence $H$ can be uncovered.) 
It is then straightforward to establish the internal energy and the entropy for each point in dynamical configuration space. 
\begin{defi}
For each configuration  $c= (\rho, H)$ in ${\cal C}({\cal H}_S)$ we define the internal energy as
\begin{eqnarray} \label{defu}
 	U(c) = \tr [\rho \, H],
\end{eqnarray} 
and the entropy as the von Neumann entropy $S(\rho)$ of $\rho$,
\begin{eqnarray} \label{defs}
 	S(\rho) = - \tr  [\rho \, \ln \rho ].
\end{eqnarray} 
\end{defi}

A central notion in thermodynamics is the canonical Gibbs state, often also referred to as \emph{thermal state} or \emph{equilibrium} state. Since a thermal state, $\rho$, at temperature $T$ is well-defined only with respect to a certain Hamiltonian, $H$, it is actually the \emph{configuration} $c = (\rho, H)$ that is thermal.
\begin{defi}
	An element $(\rho, H)\in {\cal C}({\cal H}_S)$ describes a \emph{thermal equilibrium configuration} (or briefly \emph{thermal configuration}) if  $\rho$ is a Gibbs state of the Hamiltonian, $H$, for some finite  inverse temperature $\beta >0$, i.e. 
	\begin{eqnarray} \label{eq:thermalstate}
		\rho = \frac{e^{- \beta  H }}{Z(\beta)},
	\end{eqnarray} 
with  $Z(\beta)= \tr [e^{- \beta  H}]$ being the associated partition function \footnote{Note, that in the definition we have explicitly assumed  $\beta$ to be finite. In the zero temperature limit the associated density matrix $\rho$  of the thermal configuration  $c (\beta\rightarrow \infty)$ approaches the projector on the ground state of $H$. The limiting case does however not belong to the set of the thermal configurations but to their closure. This definition ensures that the density matrices $\rho$ of the thermal configurations  $c (\beta)$ are always full rank and hence  strictly positive, i.e. the state has no zero eigenvalue.}. In the following the thermal configurations will be indicated by $c (\beta):=(\rho,H)_{\beta}$ with the subscript $\beta$ specifying the configuration's temperature.  
\end{defi} 

Thermal configurations $(\rho, H)_{\beta}$  are very special. Firstly, for a given Hamiltonian, $H$, from all possible states that have a fixed value of the internal mean energy, $U = \tr[ \rho H]$, the thermal state maximises the entropy  $S(\rho) = - \tr[\rho \ln \rho]$. In other words $c (\beta) =(\rho, H)_{\beta}$ is the most unbiased configuration one can assert to the system given only the knowledge of $U$ \cite{Jaynes1965}. Another insightful characterisation of  thermal configurations in terms of a property called \emph{complete passivity} was achieved by Lenard \cite{LENARD}, building on ideas of Pusz and Woronowicz \cite{PW}.  Complete passivity captures the intuitive notion of thermal equilibrium. A configuration $(\rho, H)$ is said to be {\em passive} if no work can be extracted from the system, i.e. $W \ge 0$ cf. Eq.~(\ref{eq:Q+W}), when subjected to any unitary transformation for a time $\tau$ generated by an arbitrary  time-dependent Hamiltonian with the sole constraint that $H(\tau)=H(0)=H$. A configuration $c=(\rho, H)$ is \emph{completely passive} if all its regularised configurations $c^{(n)}:= (\rho^{\otimes n}, H^{(n)}=\sum_{j=1}^n \, H_j)$ are passive for $n=1, 2, ...$. Here the unitary operations entering in this definition are generated by {\em arbitrary}  time-dependent Hamiltonians $H^{(n)}(t)$ that satisfy the constraint $H^{(n)}(0) = H^{(n)}(\tau) = H^{(n)}$. I.e. during $t\in ]0,\tau[$,  $H^{(n)}(t)$ is allowed to introduce any sort of  interactions between the various copies of $\rho$. It turns out that while all $c = (\rho, H)$ with commuting $\rho$ and $H$ are  passive configurations,  only thermal configurations $c (\beta)$ and the ground state are \emph{completely passive} \cite{PW,LENARD}. To stress the special role of thermal configurations graphically, they are denoted as red {circles} while all configuration that are not thermal will be called \emph{non-equilibrium} configurations and are depicted as blue squares, see Fig.~\ref{fig:config}.

\paragraph*{A note on gauge.} Given a generic state $\rho \in  \mathfrak{S} ({\cal H}_S)$  which is full rank, there always exists a Hermitian operator $H  \in {\cal L}({\cal H}_S)$ and a $\beta>0$ such that $(\rho, H)_{\beta}$ is thermal \footnote{When $\rho$ is not full rank there is no thermal configuration $c (\beta)$ whose density matrix exactly coincides with $\rho$. However, it is still possible to find Gibbs configurations whose density matrices are arbitrarily close to $\rho$. } at the inverse temperature $\beta$. In fact the problem admits infinite solutions, since there are two gauge freedoms for the choice of $H$ and $\beta$. Firstly, the zero-point of the energy scale can be chosen arbitrarily by a constant $a$. The second gauge, $b$, is the temperature itself, which sets a spacing of the energy scale. The pair $\{H, \beta\}$ is equivalent to $\{b(H+a), \beta/b\}$ in that they have the same set of thermal configurations. In particular the internal energy~(\ref{defu}) and the entropy~(\ref{defs}) of such configurations do not depend on the values of $a$ and $b$. In the following we will assume that both gauges have been chosen to some fixed values.

\section{Discrete transformations in dynamical configuration space} \label{sec:DUTs+DTTs}

Among all possible discrete transformations in dynamical configuration space ${\cal C}({\cal H}_S)$ we identify two classes that admit a clear analysis of the energetic balance and can be used as primitives for general discrete dynamical evolutions. 

\subsection{Discrete Unitary Transformations (DUTs)} 
These transformations map an initial configuration $c_i= (\rho_i, H_i)$ to a final configuration $c_f= (\rho_f,H_f)$, denoted as $c_i \stackrel{DUT}{\longrightarrow} c_f$, with the only constraint that 
\begin{eqnarray} \label{DUTsDEF}
	\rho_f = V \; \rho_i \; V^\dag \;,
\end{eqnarray} 
for some \emph{unitary} $V$. No constraint is posed on the relationship between $H_i$ and $H_f$. The definition of DUTs is inspired by continuous unitary transformations, see Eq.~(\ref{eq:unitary}). There the system is thermally isolated while evolving through the action of some external force that modifies the Hamiltonian in time, $H(t)$, and generates arbitrary unitary evolutions $V = {\cal T} e^{-i \; \int_0^\tau  d t H(t)/\hbar}$ where ${\cal T}$ indicates time ordering. For the discrete mapping $c_i \stackrel{DUT}{\longrightarrow} c_f$ no assumption is made on the time duration $\tau$ nor the specific form of $H(t)$ which realises the unitary $V$. In analogy to the continuous situation, we define the work done on the system due to a DUT identical to the total variation of the internal energy,  $\Delta U$, i.e. 
\begin{equation} \label{DUTwork}
	\begin{split} 
		W(c_i \stackrel{DUT}{\rightarrow} c_f) 
		&:=  U(c_f) - U(c_i) \\
		&= \tr[\rho_i \,  (V^{\dag} \, H_f \, V - H_i) ],  
	\end{split}
\end{equation} 
while no heat is associated with DUTs, i.e. 
\begin{equation}  \label{DUTheat} 
 	Q(c_i \stackrel{DUT}{\rightarrow} c_f)   := 0 \;. 
\end{equation} 

A special class of DUTs are the \emph{Discrete Unitary Quenches} (DUQs). Experimentally, a quench is an abrupt, instantaneous change of the system Hamiltonian which leaves the system density matrix unchanged, i.e. $V = \mathbbm{1}$, $(\rho_i, H_i) \stackrel{DUQ}{\longrightarrow} (\rho_i, H_f)$.
We also note that for full rank states $\rho_i$ a DUQ can be found that brings $(\rho_i, H_i)$ to a final configuration that is thermal,  $c_f(\beta) =({\rho_i}, \tilde{H}_f)_\beta$, with the Hamiltonian defined as $ \tilde{H}_f = - {1 \over \beta} (\ln \rho_i + \ln Z)$.

\begin{figure}[bt]
	\begin{center}
	{\includegraphics[width=\figsize\textwidth]{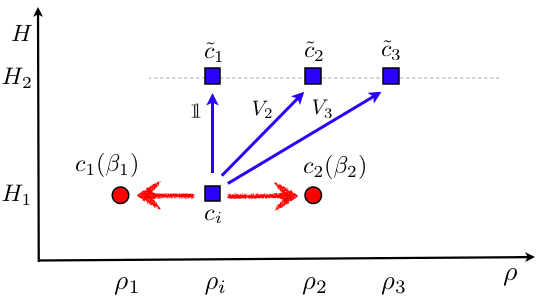}} 
\caption{Originating from the initial state $c_i = (\rho_i, H_1)$ three DUTs are shown, depicted as blue arrows, each ending at a final configuration $\tilde{c}_j = (\rho_j = V_j \, \rho_i \, V_j^{\dag}, H_2)$ for $j=1, 2, 3$. The special case where $c_i$ is transformed into $\tilde{c}_1$ is an example of a DUQ.  Also shown are two DTTs, depicted as red fuzzy arrows, originating from $c_i$ and ending at thermal configurations $c_1 (\beta_1) = (\rho_1, H_1)_{\beta_1}$ and $c_2 (\beta_2) = (\rho_2, H_1)_{\beta_3}$ where $\beta_1$ and $\beta_2$ are inverse temperatures associated with the thermal configurations. }
\label{fig:DUTs&DTTs}
\end{center}
\end{figure}

%
\medskip

DUTs will be denoted as blue arrows in the graphical representation of the configuration space, see Fig.~\ref{fig:DUTs&DTTs}.  These ``work arrows" indicate the closed nature of the transformation. The DUTs characteristic properties are summarised here:
\begin{enumerate}
	\item DUTs can start from and end at either non-thermal or thermal configurations.

	\item DUTs change the state by a unitary and no entropy change is induced by a DUT, i.e. $	\Delta S(c_i \stackrel{DUT}{\rightarrow} c_f)  = S(\rho_f) - S(\rho_i) = 0$.

	\item DUTs can be concatenated to produce another DUT.

	\item Any DUT has an inverse that is also a DUT.
\end{enumerate}

\subsection{Discrete Thermalising Transformations (DTTs)} 
DTTs are defined as those transformations which take a generic $c_i= (\rho_i, H)$  into a Gibbs state at some inverse temperature $\beta$, $ c_f (\beta) = \left( \rho_1 = {e^{-\beta H} \over Z}, H \right)_{\beta}$, without modifying the system Hamiltonian $H$. The prototypical example of a DTT is an arbitrary thermalisation process in which the system is put into a weak thermal contact with a reservoir at inverse temperature $\beta$ and left until its state becomes time-independent. Physically this is realised by the system weakly interacting with a large external environment. The requirement of a \emph{small} coupling ensures a clear definition of a local system Hamiltonian. For example, the dissipative evolution $(\rho(t), H)$ defined in Eq.~(\ref{ddf}) with the additional assumption that the Lindblad term ${\cal L}$  commutes with $H$ will for $t \rightarrow \infty$ converge to $\rho_1$. In analogy with this continuous process, we assume that the internal energy change due to a DTT is a result solely of the heat absorbed by the system
\begin{equation}   \label{DTTheat}
	\begin{split}
		Q(c_i \stackrel{DTT}{\rightarrow} c_f (\beta))  
		&: = 	U(c_f (\beta)) - U(c_i) \\
		& = \tr[H (\rho_f - \rho_i)],
	\end{split}
\end{equation} 
while the work of a DTT vanishes, 
\begin{equation}  \label{DTTwork}
	W(c_i \stackrel{DTT}{\rightarrow} c_f (\beta)) := 0.
\end{equation}
This non-trivial expression of the heat is exactly of the form that we expected from the bounds on the entropy in Eq.~(\ref{eq:entropy}), and it will be the basis for deriving a general second law for discrete quantum trajectories in the next section. 

\medskip

DTTs will be denoted as horizontal red arrows in the graphical representation, see Fig.~\ref{fig:DUTs&DTTs}. The fuzziness of these ``heat arrows"  indicates the open nature of the transformation. The characteristic properties of DTTs are summarised here: 

\begin{enumerate}

\item DTTs always end in thermal configurations.

\item DTTs do not change the Hamiltonian.

\item The entropy change associated with a DTT  is in general nonzero, i.e.  $\Delta S(c_i \stackrel{DTT}{\rightarrow} c_f(\beta))  = S(\rho_f) - S(\rho_i) \not = 0.$

\item DTTs can be concatenated to produce another DTT.

\item The inverse of a DTTs is in general not a DTT. Only if the initial configuration $c_i$ was already thermal, can the action of a DTT be reversed by another DTT.

\end{enumerate} 

\section{Heat and Clausius inequality for discrete thermodynamic processes} \label{sec:2ndlaw}

Having identified two fundamental process primitives in configuration space, we now focus on more complex discrete trajectories. These can start from equilibrium or non-equilibrium configurations, however, we restrict ourselves to discrete trajectories that can be obtained by concatenating DUT and DTTs. Within this scenario we will be able to formulate a general second law for discrete quantum processes, that does not require detailed knowledge of the continuous state and local Hamiltonian evolution.

\subsection{Single DUT+DTT transformations}

Let us begin with the simplest non trivial discrete transformation which can be used to connect two equilibrium configurations. 

\subsubsection{Equilibrium to equilibrium processes}

\begin{figure}[t]
	\begin{center}
	{\includegraphics[width=\figsize\textwidth]{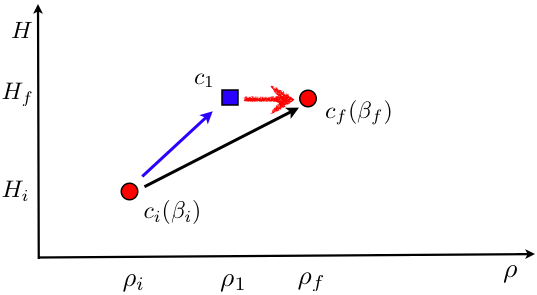}}
\caption{Trajectory connecting two Gibbs configurations,  $c_i (\beta_i) = (\rho_i, H_i)_{\beta_i} \rightarrow c_f (\beta_f) = (\rho_f, H_f)_{\beta_f}$ (black arrow) and discrete decomposition into a DUT, $c_i (\beta_i) \rightarrow c_1$ (blue arrow), to an intermediate point $c_1 = (\rho_1, H_f)$ followed by a DTT, $c_1 \rightarrow c_f (\beta_f)$ (red arrow).}
\label{fig:DUT+DTTtraj}
\end{center}
\end{figure}

We consider a trajectory that starts from a thermal configuration $c_i (\beta_i) = (\rho_i, H_i)_{\beta_i}$ and ends at a final thermal configuration $c_f (\beta_f) = (\rho_f, H_f)_{\beta_f}$ 
via the action of a single DUT followed by a DTT. The heat of the discrete process can then be determined as the sum of the heats of each component, for both of which the heat is a well-defined quantity. The DUT first unitarily rotates the input density matrix to $\rho_1 = V \, \rho_i \, V^{\dag}$ while the Hamiltonian changes from $H_i$ to $H_f$, ending in an intermediate (not necessarily thermal) configuration $c_1 = (\rho_1, H_f)$. A DTT follows that brings $c_1$ to $c_f (\beta_f)$, resulting in the discrete overall trajectory
\begin{eqnarray} \label{eq:DUT+DTT}
	c_i (\beta_i)   \stackrel{DUT}{\longrightarrow}  
	  c_1  \stackrel{DTT}{\longrightarrow}  c_f (\beta_f), 
\end{eqnarray}
shown in Fig.~\ref{fig:DUT+DTTtraj}. 
While heat of process (\ref{eq:DUT+DTT}) is only exchanged during the DTT,  the amount of exchanged heat depends on the DUTs unitary $V$
\begin{eqnarray} \label{eq:heatDUT+DTT}
	Q(c_1 \stackrel{DTT}{\rightarrow} c_f (\beta_f))  
	=  \tr [(\rho_f - V \, \rho_i \, V^{\dag}) \; H_f ].
\end{eqnarray} 
Clearly, the value of the heat depends on the choice of the unitary $V$ with the maximum and minimum heat given by
\begin{equation}
	\begin{split}
	Q_{\max} 
	&=   \sum_{k=1}^N H_f(k) \Big(\frac{e^{- \beta_f H_f(k)}}{Z_f} -  \frac{e^{- \beta_i H_i(k)}}{Z_i} \Big),\\
   	Q_{\min} 
	&=   \sum_{k=1}^N  H_f(k) \Big(\frac{e^{- \beta_f H_f(k)}}{Z_f} -  \frac{e^{- \beta_i H_i(N-k+1)}}{Z_i}\Big),
	\end{split}
\end{equation} 
where $\{H_f(k)\}_k$ and $\{H_i(k)\}_k$ are the eigenvalues of $H_f$ and $H_i$ ordered in decreasing order and $Z_{i,f}$ the partition functions of the initial and final configuration. The derivation of this expression and the corresponding unitaries $V_{\min}$ and $V_{\max}$, are given in Appendix \ref{sec:min&max}. 

However, for any possible choice of the DUT connecting $c_i (\beta_i)$ to the intermediate step $c_1$, i.e. for any unitary transformation $V$, a second law can be established by linking the heat and the entropy change. The entropy change is bounded according to Eq.~(\ref{eq:entropybounds}), by
\begin{eqnarray}   \label{eq:s-bound}
	\begin{aligned}	
		\Delta S(\rho_i, \rho_f) 
		&=   S(\rho_f) - S(\rho_1)   \\
		&\geqslant  - \tr [ (\rho_f -\rho_1) \ln\rho_f]  \\
		&= \beta_f \;  \tr [(\rho_f - V \, \rho_i \, V^{\dag}) \; H_f ] . 
	\end{aligned}
\end{eqnarray}
implying
\begin{equation} \label{eq:gg1113}
 	\Delta S(\rho_i, \rho_f) 
	\geqslant  \beta_f \; 	Q(c_i {\rightarrow} c_f) .
\end{equation} 
Thus the process (\ref{eq:DUT+DTT}) obeys a Clausius-type inequality, cf. (\ref{eq:Clausius}), which states that the heat absorbed by the system is upper bounded by the entropy change.

\subsubsection{Non-equilibrium to non-equilibrium processes}

We now turn to discrete non-equilibrium processes for which establishing the Clausius inequality in the continuous case has only recently been addressed \cite{Esposito2011}. 
In our approach this can be done by observing that 
 given two generic configurations $c_i= (\rho_i, H_i)$ and $c_f=(\rho_f, H_f)$ in ${\cal C}({\cal H}_S)$, it always possible to connect them 
via a discrete trajectory composed by three primitive steps which differs from the one given in Eq.~(\ref{eq:DUT+DTT}) only by a final DUQ transformation. Specifically we can write 
\begin{eqnarray}  \label{real1333} 
	{\cal T} := c_i  \stackrel{DUT}{\longrightarrow} {c}_1 \stackrel{DTT}{\longrightarrow} c_2 (\beta_2) \stackrel{DUQ}{\longrightarrow} c_f ,
\end{eqnarray}
with intermediate configurations ${c}_1=(V \, \rho_i \, V^{\dag}, H_1)$ and $c_2 (\beta_2) = (\rho_f, H_1)_{\beta_2}$. 
Notice that the first and the last step of~(\ref{real1333}) do not alter the entropy of the system, nor contribute 
to the heat exchange since they are DUTs. This implies the identity $\Delta S(\rho_i, \rho_f)  = \Delta S(\rho_1,\rho_2)$ and 
allows us to identify  the heat associated with ${\cal T}$  with the quantity  
\begin{equation}
	Q({\cal T}) = Q({ c}_1\stackrel{DTT}{\longrightarrow} {c}_2 (\beta_2)) = \tr [(\rho_f-  V \, \rho_i \, V^{\dag}) H_1] .
\end{equation}
The lower bound of Eq.~(\ref{eq:entropybounds}) can then be used to establish a Clausius-type inequality for the discrete transformation~(\ref{real1333}), i.e. 
\begin{eqnarray} \label{eq:gg111333}
	\Delta S(\rho_i, \rho_f)  &=& \Delta S(\rho_1,\rho_2) \\ \nonumber 
	&\geqslant& - \tr [(\rho_f-  V \, \rho_i \, V^{\dag}) \ln \rho_f] 
	= \beta_2 \; Q({\cal T}),
\end{eqnarray} 
where $\beta_2$ is the temperature of the intermediate configuration $c_2(\beta_2)$.

\subsection{Sequences of DUT+DTTs}  \label{sec:multiple}

\begin{figure}[t]
	\flushleft 
	{\bf \large a} \\
	{\includegraphics[width=\figsize\textwidth]{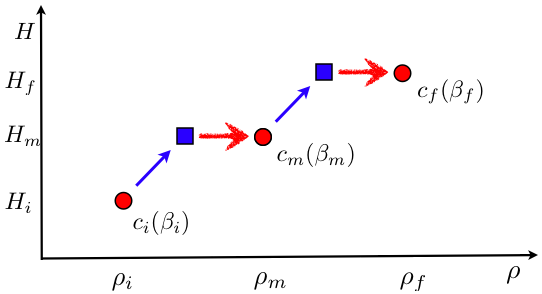}} \\
	{\bf \large b} \\
	{\includegraphics[width=\figsize\textwidth]{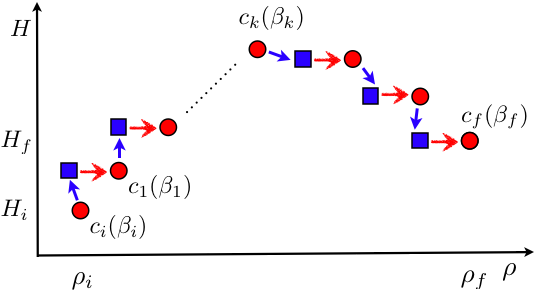}}
\caption{ Panel {\bf a} shows the realisation of a general transformation $c_i (\beta_i) \rightarrow c_f (\beta_f)$ as a sequence of two DUT+DTT transformations, via an intermediate thermal configuration, $ c_m (\beta_m) = (\rho_m, H_m)_{\beta_m}$. Panel {\bf b}  shows the realisation of a general transformation $c_i (\beta_i) \rightarrow c_f (\beta_f)$ through a sequence of DUT+DTT transformations, via many intermediate thermal configurations, $c_k (\beta_k) = (\rho_k, H_k)_{\beta_k}$.}
\label{fig:multiple}
\end{figure}

The trajectories defined in Eq.~(\ref{eq:DUT+DTT}) and Eq.~(\ref{real1333}) 
are just specific choices of  discrete trajectories connecting two configurations $c_i $ and $c_f $. We will now show that a Clausius inequality, e.g. inequalities of the type (\ref{eq:gg1113}),  holds for general discrete processes as long as they can be decomposed in a \emph{sequence} of DUT+DTTs steps. 
To show this, we first consider the discrete trajectory, $\gamma$, pictured in panel {\bf a} of Fig.~\ref{fig:multiple} where $c_i (\beta_i)= (\rho_i, H_i)_{\beta_i}$ is transformed into $c_f (\beta_f) =(\rho_f, H_f)_{\beta_f}$ via two DUT+DTT transformations and a third thermal configuration $c_m (\beta_m)=(\rho_m, H_m)_{\beta_m}$, i.e. 
\begin{eqnarray} \label{real222}
	\gamma:= c_i (\beta_i)  \, \stackrel{DUT+DTT}{\longrightarrow}  \,
	c_m (\beta_m) 	\, \stackrel{DUT+DTT}\longrightarrow  \, 
	c_f (\beta_f).	
\end{eqnarray}
In this scenario the following inequality for the entropy holds,
\begin{equation} 	 \label{hhVV} 
	\begin{split}
		\Delta S(\rho_i, \rho_f) &=  \Delta S(\rho_i, \rho_m)+  \Delta S(\rho_m, \rho_f) \\
		&\geqslant \beta_m \;  Q_m  +  \beta_f \; Q_f,
	\end{split}
\end{equation}
where $Q_m = Q(c_i (\beta_i) \stackrel{DUT+DTT}{\longrightarrow} c_m (\beta_m))$ and $Q_f =Q(c_m (\beta_m) \stackrel{DUT+DTT}{\longrightarrow} c_f (\beta_f)) $, and where Eq.~(\ref{eq:gg1113}) was used for the two DUT+DTT transformations.

Inequality~(\ref{hhVV})  can immediately be generalised to an arbitrary number of intermediate  DUT+DTT  steps connecting $c_i(\beta_i)$ to $c_f(\beta_f)$. Specifically, consider a generic discrete trajectory  ${\cal T}$ composed of $N$ consecutive  DUT+DTT steps that pass through the thermal configurations $\{ c_i(\beta_i), c_{1}(\beta_{1}), c_{2}(\beta_{2}), \cdots, c_{N-1}(\beta_{N-1}), c_f(\beta_f) \}$ as shown in panel {\bf b} in Fig.~\ref{fig:multiple}. Then by expressing the total entropy increment  $\Delta S(\rho_i, \rho_f)$ as a sum of terms $\Delta S(\rho_{k}, \rho_{k+1})$ associated with the various steps of ${\cal T}$  and  applying Eq.~(\ref{eq:gg1113}) to each one of them, the Clausius inequality becomes 
\begin{equation}  \label{yyy1} 
	\begin{split}
		\Delta S(\rho_i, \rho_f)  \geqslant \sum_{k=0}^{N-1}  \beta_{k+1} \, Q(c_{k} (\beta_{k}) \stackrel{DUT+DTT}{\longrightarrow} c_{k+1} (\beta_{k+1})).
	\end{split}
\end{equation}
The generality of this derivation implies that sequences of discrete unitary and discrete thermalising transformations always fulfil a Clausius type equation. 

To formulate this as a lemma, we introduce a useful discrete process quantity, $\Lambda$, for a DUT+DTT sequence,
\begin{eqnarray} \label{eq:Lambdapiece}	
	\begin{split}	
		& \Lambda (c_k (\beta_k) \stackrel{DUT+DTT}{\longrightarrow} c_{k+1} (\beta_{k+1})) \\
		&:=	 \beta_{k+1} \tr[(\rho_{k+1}-V_k \, \rho_k \, V_k^{\dag}) \,H_{k+1} ], \\
		&= 	\beta_{k+1} \, Q(c_{k} (\beta_{k}) \stackrel{DUT+DTT}{\longrightarrow} c_{k+1} (\beta_{k+1})).
	\end{split}
\end{eqnarray}
The quantity $\Lambda$ is the discrete analog to the continuous expression $\int {\de Q \over T}$. For the trajectory $\cal{T}$ the overall $\Lambda({\cal T})$ is obtained by summing over the $\Lambda$ contributions of the various steps, i.e. 
\begin{eqnarray} 	 \label{functionalS}
	\Lambda({\cal T}) := \sum_{k=0}^{N-1} \Lambda (c_k (\beta_k) \stackrel{DUT+DTT}{\longrightarrow} c_{k+1} (\beta_{k+1})),
\end{eqnarray}
with $k=0$ and $k=N$ corresponding to the initial and final configurations, $i$ and $f$, respectively. Then the lower entropy bound, Eq.~(\ref{yyy1}), can be neatly expressed:

\begin{lemma}
Any trajectory ${\cal T}$ made of sequences of DUT+DTTs fulfils a Clausius inequality of the form
\begin{eqnarray} \label{eq:SL}
	\Delta S(\rho_i, \rho_f) \geqslant     \Lambda({\cal T}).
\end{eqnarray}
\end{lemma}

While in general a single DUT+DTT process cannot saturate the equality, see Appendix \ref{sec:min&max}, we will now show that augmenting intermediate steps will always increase the r.h.s. of Eq.~(\ref{yyy1}). Moreover, we find that in the limit of infinitely long sequences the asymptotic saturation of the inequality~(\ref{yyy1}) is always possible.

\subsection{Saturating the Clausius bound} \label{sec:sat}  

\begin{figure}[t]
	\begin{center}
	{\includegraphics[width=\figsize\textwidth]{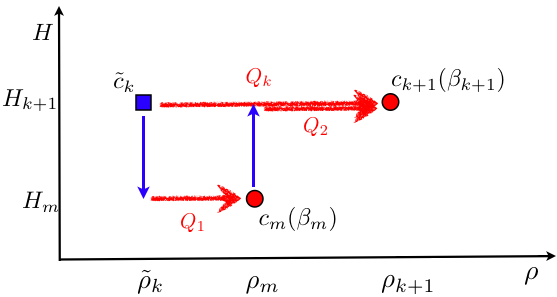}}
\caption{Discrete trajectory connecting $\tilde{c}_{k} \rightarrow c_{k+1} (\beta_{k+1})$ by a single DTT with heat $Q_k$, and by a sequence of two DUQ+DTT transformations via an intermediate thermal configurations, $c_m (\beta_m) = (\rho_m, H_m)_{\beta_m}$. The two DTTs have heats $Q_1$ and $Q_2$, respectively.}
\label{fig:augmenting}
\end{center}
\end{figure}

To show that the entropy bound Eq.~(\ref{yyy1}) can be saturated we construct a class of trajectories ${\cal T}'$ from a generic DUT+DTT trajectory ${\cal T}$, as depicted in panel {\bf b} of Fig~\ref{fig:multiple}, for which the functional $\Lambda({\cal T}')$ is always larger than $\Lambda({\cal T})$. One class of trajectories ${\cal T}'$ is the trajectory identical to ${\cal T}$ however with the step $\tilde{c}_{k} \stackrel{DTT}{\longrightarrow} c_{k+1} (\beta_{k+1})$ replaced with the sequence  $\tilde{c}_{k} \stackrel{DUQ+DTT} {\longrightarrow} c_m (\beta_m) \stackrel{DUQ+DTT}{\longrightarrow} c_{k+1} (\beta_{k+1})$  as shown in Fig.~\ref{fig:augmenting}, where $c_m (\beta_m)= (\rho_m , H_m)_{\beta_m}$ has an intermediate density matrix
\begin{eqnarray} \label{middlepoint}
 	\rho_m = p \; \tilde{\rho}_{k} + (1-p)\;  \rho_{k+1}, 
 \end{eqnarray}
with mixing probability $p\in ]0,1[$ \footnote{$c_m (\beta_m)$ is a well-defined thermal configuration as its state is by construction full rank. }. The increment on the r.h.s. of (\ref{eq:SL}) for the new trajectory $\cal{T}'$, $\Delta({\cal T}',{\cal T})  = \Lambda ({\cal T}') -  \Lambda ({\cal T}) = \beta_{m} Q_1  + \beta_{k+1} Q_2 - \beta_{k+1} Q_k$, see Fig.~\ref{fig:augmenting}, is then \emph{strictly positive} for any $p$
\begin{widetext}
\begin{equation} 
	\begin{aligned}
	\Delta({\cal T}',{\cal T}) 
	&= -\tr [ \ln \rho_m \, (\rho_m - \tilde{\rho}_{k})] -\tr [ \ln \rho_{k+1} \, (\rho_{k+1} - \rho_{m})]  + \tr [ \ln \rho_{k+1} \, (\rho_{k+1} - \tilde{\rho}_{k})] \nonumber\\
	& >  - p S( \tilde{\rho}_{k} \| \tilde{\rho}_{k}) - (1-p) \; S( \rho_{k+1}  \| \tilde{\rho}_{k}) - p S(\rho_{k+1} \|  \tilde{\rho}_{k} )  -  (1-p)\;  S(\rho_{k+1} \| \rho_{k+1}) + S(\rho_{k+1} \| \tilde{\rho}_{k}) \\
	&= 0,
		\end{aligned}
\end{equation} 
\end{widetext}
where we have assumed $\tilde{\rho}_k \not = \rho_{k+1}$, and used Eq.~(\ref{middlepoint}) and the joint convexity of the relative entropy~\cite{PETZ}, 
\begin{equation}
	\begin{split}
		S(\rho_1 \| p \tilde{\rho}_k + (1-p) \rho_{k+1})  &\leqslant p S(\rho_1 \| \tilde{\rho}_k ) +  (1-p) S(\rho_1 \| \rho_{k+1})\\
		S(p \tilde{\rho}_k + (1-p) \rho_{k+1} \| \rho_1)  &\leqslant p S(\tilde{\rho}_k \| \rho_1 ) +  (1-p) S(\rho_{k+1} \| \rho_1)
	\end{split}
\end{equation}
with equality \emph{iff} $\rho_{k+1} = \rho_1 = \tilde{\rho}_k$. We summarise this result in the following Lemma:

\begin{lemma}
Adding intermediate thermal configurations $c_m(\beta_m)$ (see Eq.~\ref{middlepoint}) to any trajectory ${\cal T}$ results in a new trajectory ${\cal T'}$ with increased $\Lambda$, i.e.
\begin{eqnarray} 
	\Delta({\cal T}',{\cal T})  = \Lambda ({\cal T}') -  \Lambda ({\cal T}) > 0.
\end{eqnarray}
\end{lemma}


Having confirmed that it is possible for any given discrete trajectory to  introduce intermediate steps such that the r.h.s. of Eq.~(\ref{eq:SL}) increases, the task is now to show that the entropy bounds can be saturated by  reiterating the procedure. The proof relies on lower bounding $\Lambda$ and showing that the bound converges to the upper bound on $\Lambda$, Eq.~(\ref{eq:SL}), in the limit of infinite steps. The detailed derivation is given in Appendix \ref{sec:satproof} proving the following theorem:
\begin{theo} \label{lem:Clausiussat}
	Let ${\cal T}$ be a discrete trajectory  connecting the initial Gibbs configuration $c_i (\beta_i)=(\rho_i,H_i)_{\beta_i}$ to the final Gibbs configuration $c_f (\beta_f)=(\rho_f,H_f)_{\beta_f}$ via a sequence of $N$ concatenated   DUT+DTT steps $c_{k} (\beta_k) \stackrel{DUT+DTT}{\longrightarrow}  c_{k+1} (\beta_{k+1})$ connecting  the thermal configurations ${\cal T} = \{ c_i (\beta_i) = c_0 (\beta_0), c_1 (\beta_1), ... c_N (\beta_N) = c_f (\beta_f)\}$ as in panel {\bf b} in Fig.~\ref{fig:multiple}. Then a sequence of trajectories ${\cal T}'_n$ exists, obtained from ${\cal T}$ by adding $n$ intermediate thermal steps, which saturates the Clausius  bound~(\ref{yyy1}) in the asymptotic limit, i.e.  
\begin{equation}\label{thesisnew} 
	\Delta S(\rho_i,\rho_f) = \lim_{n\rightarrow \infty} \Lambda( {\cal T}'_n).
\end{equation}
\end{theo}

\subsection{Approximation of continuous processes by discrete processes}

In the introduction we have seen that for continuous processes where consistent definitions of $\rho(t)$ and of the local Hamiltonian $H(t)$ can be assigned for all $t$, Eq.~(\ref{eq:Q}) defines the heat absorbed by the system. We have already discussed the difficulties of knowing $\rho(t)$ and identifying a proper local Hamiltonian $H(t)$ for the system. However, in what follows we will assume that some ``valid" continuous trajectory $c(t) = (\rho(t), H(t)) \in {\cal C}({\cal H}_s)$ is given for which Eqs.~(\ref{eq:Q}) and~(\ref{eq:Q+W}) apply. We now wish to identify a discrete trajectory that connects the same initial and final configurations as the continuous trajectory through a sequence of DUTs and DTTs, and which approximates the continuous heat. The analysis leads to the following theorem:

\begin{theo}
For a continuous process between two configurations $c_0$ and $c_{\tau}$ that obeys the Clausius inequality, a discrete trajectory $\Gamma'$ exists that connects the same configurations and has exactly the same heat as the continuous process.
\end{theo}

\emph{Proof:} Consider an infinitesimal heat increment along the continuous trajectory,
\begin{eqnarray} \label{quantity}
	{\de}Q(t) = \tr [ (\rho(t) - \rho(t-{d}t)) H(t) ],
\end{eqnarray}
with $\rho(t)$ and $\rho(t - {d}t)$ being the density matrices of two infinitesimally separated configurations on the trajectory. Define the initial and final configuration for a discrete trajectory to be  $c_i = (\rho(t - {d}t), H(t - {d}t))=: (\rho_1, H_i)$ and $c_f = (\rho(t), H(t)):= (\rho_2, \tilde{H}_2)$. To compare the continuous heat~(\ref{quantity}) with a discrete heat we need to identify a discrete trajectory, $\gamma$, connecting the same initial and final configuration as the continuous trajectory. 

\begin{figure}[t]
	\begin{center}
	{\includegraphics[width=\figsize\textwidth]{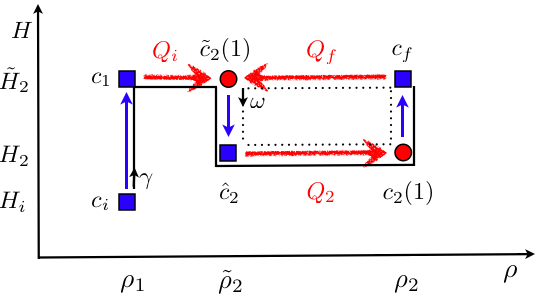}}
\caption{The figure shows the  initial and final non-equilibrium configurations, $c_i$ and $c_f$, of the continuous process. A discrete trajectory, $\gamma$, that connects $c_i$ and $c_f$ is also indicated (solid).  Also shown is the closed loop discrete trajectory $\omega$ (dotted). The two thermal configurations $\tilde{c}_2 (1) = (\tilde{\rho}_2, \tilde{H}_2)_{\beta =1}$ and $c_2 (1) = (\rho_2, H_2)_{\beta =1}$ are both chosen with inverse temperature $\beta =1$, as indicated in the circles.
}
\label{fig:triangle}
\end{center}
\end{figure}

One example is the sequence $\gamma$ shown as a solid line in Fig.~\ref{fig:triangle}, 
\begin{equation}\label{trj}
	\gamma:=  c_i \stackrel{DUQ}{\rightarrow} c_1 \stackrel{DTT}{\rightarrow} \tilde{c}_2 (1) \stackrel{DUQ}{\rightarrow} \hat{{ c}}_2 \stackrel{DTT}{\rightarrow}  c_2 (1) \stackrel{DUQ}{\rightarrow}  c_f,
\end{equation}
where $\tilde{c}_2 (1)= (\tilde{\rho}_2, \tilde{H}_2)_{\beta=1}$ and $c_2 (1) = (\rho_2, H_2)_{\beta =1}$ are equilibrium configurations \footnote{A proper definition of $c_2 (1)$ requires $\rho_f$ to be full rank. If this is not the case one can still define a trajectory (\ref{trj}) with $c_f$ replaced by a full rank  configuration which can be chosen to be arbitrarily close to $c_f$.} 
and $\hat{c}_2 = (\tilde{\rho}_2, H_2)$ is a non-equilibrium configuration. The inverse temperatures of the equilibrium configurations $\tilde{c}_2 (1)$ and $c_2 (1)$ are both chosen $\beta = 1$. Heat $Q_i$, 
\begin{equation} \label{defq1}
	\begin{split}
		Q_i :&= Q(c_i \stackrel{DUQ+DTT}{\longrightarrow} \tilde{c}_2 (1)) 
		= Q (c_1 \stackrel{DTT}{\longrightarrow} \tilde{c}_2 (1)) \\
		&= - \tr [(\tilde{\rho}_2 - \rho_1 ) \ln \tilde{\rho}_2],
	\end{split}
\end{equation}
is exchanged when passing from $c_0\rightarrow \tilde{c}_2 (1)$ via a DUQ+DTT through the intermediate configuration $c_1$, see Fig.~\ref{fig:triangle}. Heat $Q_2$,
\begin{equation}
	Q_2 := Q (\hat{c}_2 \stackrel{DTT}{\longrightarrow} c_2 (1) ) = - \tr [(\rho_2 - \tilde{\rho}_2) \ln {\rho}_2],
\end{equation}
is exchanged when passing from $\hat{c}_2 \rightarrow c_2 (1)$. Therefore 
\begin{eqnarray} \label{DEFQGAMMA}
	Q(\gamma) = {}Q_i + {}Q_2\;.
\end{eqnarray}

On the other hand, the continuous heat increment $\de Q(t)$  can be decomposed into two heat contributions,
\begin{equation} \label{defdq}
	\begin{aligned}
		{\de}Q(t) &=  \tr [ (\rho(t) - \rho(t- {d}t)) H(t) ] = \tr [(\rho_2 - \rho_1) \tilde{H}_2] \\
		&=  \tr [ ({\rho}_2 -\tilde{\rho}_2) \tilde{H}_2 ] +  \tr [ (\tilde{\rho}_2 -\rho_1 ) \tilde{H}_2 ]   \\
		&= - \tr [ ({\rho}_2 -\tilde{\rho}_2) \ln \tilde{\rho}_2  ] - \tr [ (\tilde{\rho}_2 -\rho_1 ) \ln \tilde{\rho}_2]  \\
		&= - {}Q_f + {}Q_i.
	\end{aligned}
\end{equation}
where $Q_f$ would be the heat absorbed by the system if it passed from $c_f$ to $\tilde{c}_2 (1)$ via a  DTT, i.e. 
\begin{equation} \label{defq2}
	Q_f := Q ({c}_f \stackrel{DTT}{\longrightarrow} \tilde{c}_2 (1))
		= - \tr [(\tilde{\rho}_2 -\rho_2) \ln \tilde{\rho}_2] .
\end{equation}
To compare the continuous heat (\ref{defdq}) with the discrete heat (\ref{DEFQGAMMA}), we introduce a second discrete trajectory $\omega$. This is a closed loop sequence of DUQ and DTT transformations, see Fig.~\ref{fig:triangle},
\begin{eqnarray}\label{trjqq} \label{omega}
	\omega := \tilde{c}_2 (1) \stackrel{DUQ}{\rightarrow} \hat{c}_2 \stackrel{DTT}{\rightarrow}  c_2 (1) \stackrel{DUQ}{\rightarrow}  c_f  \stackrel{DTT}{\rightarrow} \tilde{c}_2 (1).
\end{eqnarray}
In trajectory $\omega$ heat is exchanged from $\hat{c}_2 \to c_2 (1)$ and from $c_f \to \tilde{c}_2 (1)$, so $Q(\omega) = Q_2 + Q_f$. 
As discussed in previous sections the discrete heat always obeys the Clausius inequality (\ref{eq:gg1113}), and with $\beta=1$ for both steps this implies 
\begin{eqnarray} \label{eq:loop}
	0 = \Delta S(\tilde{\rho}_2, \tilde{\rho}_2) \geqslant Q_2 + Q_f\;
\end{eqnarray}
for trajectory $\omega$. Using this in Eq.~(\ref{DEFQGAMMA}) we find that the heat associated with the discrete trajectory $\gamma$ from $c_i$ to $c_f$ is a \emph{lower bound} to the infinitesimal continuous heat for the same initial and final configuration, (\ref{quantity}), i.e. 
\begin{eqnarray}
	Q(\gamma) = Q_i + Q_2\; \leqslant Q_i - Q_f = {\de}Q (t)\;.
\end{eqnarray}
This can immediately be extended to the full continuous process: For any arbitrary continuous process between $c_0$ and $c_{\tau}$ there is always a discrete trajectory $\Gamma$ between the same two configurations that has a lower heat than the continuous heat, Eq.~(\ref{eq:Q}). Moreover, by augmenting intermediate steps in the discrete trajectory $\Gamma$, resulting in the trajectory $\Gamma'$ that passes through an infinite sequence of points $c(t)$, it is possible to increase the associated heat, as shown in Sec.~\ref{sec:multiple}. From Theorem \ref{lem:Clausiussat} follows that if the continuous trajectory fulfils the Clausius inequality, then a discrete trajectory connecting the same initial and final configuration can be found that has the same heat as the continuous trajectory. $\blacksquare$

\section{Thermal efficiency}  \label{sec:efficiency}

\begin{figure}[t]
	\begin{center}
	{\includegraphics[width=\figsize\textwidth]{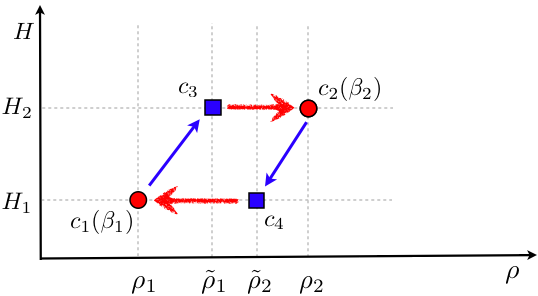}}
\caption{Representation of the cycle of Eq.~(\ref{LOOP}). }
\label{fig:cycle}
\end{center}
\end{figure}

The last piece in our analysis of the energy balance in discrete quantum processes is to determine the efficiency of a discrete cyclic process, such as the one depicted in Fig.~\ref{fig:cycle} where $c_1 (\beta_1) = (\rho_1, H_1)_{\beta_1}$, $c_2 (\beta_2)= (\rho_2, H_2)_{\beta_2}$ are equilibrium configurations, while $c_3 = (\tilde{\rho}_1, H_2)$, and  $c_4 = (\tilde{\rho}_2, H_1)$ are not. This results in the following Lemma:

\begin{lemma} The thermal efficiency of the discrete cycle depicted in Fig.~\ref{fig:cycle} is bounded by the Carnot efficiency,
\begin{eqnarray} 
	\eta 	& \leqslant  1 - \frac{T_1}{T_2},
\end{eqnarray}
and the optimal efficiency is achievable.
\end{lemma}

{\it Proof:} For the overall loop,
\begin{eqnarray} \label{LOOP}
	c_1 (\beta_1) \stackrel{DUT}{\rightarrow} c_3  \stackrel{DTT}{\rightarrow} c_2 (\beta_2)   \stackrel{DUT}{\rightarrow} c_4 \stackrel{DTT}{\rightarrow} c_1 (\beta_1) \;,
\end{eqnarray}
the entropy change is zero. However, the entropy change nevertheless bounds the heat of the two heat-producing DTT processes, $c_3\rightarrow c_2 (\beta_2)$ and $c_4\rightarrow c_1 (\beta_1) $,
\begin{equation} \label{eq:heatbound} 
	\begin{split}
		\Delta S= 0 
		& \geqslant  \beta_2 \, Q(c_3\rightarrow c_2) + \beta_1 \, Q(c_4\rightarrow c_1).
	\end{split}
\end{equation}  

This implies that at least one of the two heats must be negative. Let us assume for instance that the heat exchanged with the thermal reservoir at temperature $T_2 = 1/(k_B \beta_2)$ is positive, $Q(c_3\rightarrow c_2) >  0$,  while the other heat is negative $ Q(c_4\rightarrow c_1) < 0$ (other scenarios can be treated analogously, see below). The total heat absorbed per cycle is 
\begin{eqnarray} 
	Q(c_1\rightarrow c_1) = Q(c_3\rightarrow c_2) + Q(c_4\rightarrow c_1),
\end{eqnarray}
with the energy balance implying that the overall absorbed heat must be equal to the negative work done on the system during the cycle, 
\begin{eqnarray} 
	0 = \Delta U(c_1\rightarrow c_1) = Q(c_1\rightarrow c_1) + W(c_1\rightarrow c_1).
\end{eqnarray}
The thermal efficiency is defined as the ratio between the work performed and the heat absorbed, leading to
\begin{equation} \label{eta}
	\begin{split}
		\eta &= \frac{-W(c_1\rightarrow c_1) }{Q(c_3\rightarrow c_2)} 
			= \frac{Q(c_1\rightarrow c_1) }{Q(c_3\rightarrow c_2)} \\
		& = \frac{Q(c_3\rightarrow c_2)+ Q(c_4\rightarrow c_1)}{Q(c_3\rightarrow c_2)}  \\
		& \leqslant 1- \frac{\beta_2}{\beta_1} = 1 - \frac{T_1}{T_2} , 
		\end{split}
\end{equation} 
where we used Eq.~(\ref{eq:heatbound}). If $T_1\leqslant T_2$ the system absorbs heat from a higher temperature bath and gives heat to a lower temperature bath. The efficiency $\eta$ is then positive and smaller than unity with the optimal efficiency reproducing the Carnot efficiency. The optimal efficiency can be reached by augmenting the discrete trajectory to saturate the equality in the Clausius-inequality, see Theorem  \ref{lem:Clausiussat}. $\blacksquare$

\medskip

{\it Remark:} If instead  $T_2\geqslant T_1$, i.e. heat is absorbed from a lower temperature bath and given to one at a higher temperature, the system operates as a refrigerator. In this case the total work \emph{absorbed} by the system, $ W(c_1\rightarrow c_1) = Q(c_3\rightarrow c_2) (T_1/T_2 -1)$, is positive. The efficiency of the process can be measured by the coefficient of performance, COP$_{cooling}$, defined as the ratio between the heat absorbed from the cold reservoir $T_2$ (i.e. $Q(c_3\rightarrow c_2)$) and the total work done on the system 
\begin{eqnarray}
	\mbox{COP}_{cooling} = \frac{Q(c_3\rightarrow c_2)}{W(c_1\rightarrow c_1)} \leqslant \frac{T_2}{T_1-T_2}\;,  \label{cooling}
\end{eqnarray} 
which again is always smaller than one. Finally, if the signs for the heats $Q(c_3\rightarrow c_2)$ and $Q(c_4\rightarrow c_1)$ are interchanged the above argument still holds with Eqs.~(\ref{eta}) and (\ref{cooling}) being replaced by 
the inequalities $\eta \le 1 - {T_2 \over T_1}$ and $\mbox{COP}_{cooling} \leqslant \frac{T_1}{T_2-T_1}$, respectively.

\section{Conclusions}  \label{sec:Conc}

The early development of thermodynamics culminating in the formulation of the second law also gave birth to a new quantity, the entropy, whose physical meaning was at first opaque. Only later was its meaning elucidated by the works of Boltzmann and others. In this paper we proposed to turn the original argument around and use the well-established notion of entropy that characterises the information content in a (quantum) state to motivate the definition of a notion of heat for discrete quantum processes. The approach circumvents the large cluster of problems surrounding the idea of a unique definition of heat and work in processes where (i) the Hamiltonian of the system is not well-defined due to the open nature of the system and (ii) there are fundamental limitations on the notion of trajectories where full knowledge is only given at discrete points in time when a measurement with a specific Hamiltonian occurred. 

By introducing thermodynamic configurations, identifying two primitives for discrete processes, DUTs and DTTs, and defining heat to pertain only to DTTs we were able to uncover a general second law valid for any discrete process consisting of sequences of DUT+DTTs between both, equilibrium and non-equilibrium configurations. Moreover, we showed that an infinite sequence of DUT+DTT processes exists that saturates the Clausius inequality. In other words, saturation occurs when a discrete trajectory is mapped out into a continuous one by a sequence of measurements that are infinitely close together. This provides a link between \emph{reversibility} - here the reversibility of a discrete process - and the equality in the second law for discrete processes, reminiscent of the Clausius' statement of equality for thermodynamic reversible continuous processes, Eq.~(\ref{eq:reversible}). On the other hand, we also showed that for any continuous process between two configurations that obeys the Clausius inequality, there exists a discrete process between the same configurations with the same heat. Finally, we showed that for the discrete version of a thermodynamic cycle, formed by a discrete trajectory passing through four configurations Carnot's efficiency is recovered. 

The strength of our approach is to give meaning to heat and work, reversibility and efficiency following from just a few sensible and simple definitions. In some respect 
this is analogous  to the axiomatic approach to thermodynamics first developed by Carath\'{e}odory~\cite{CARA}.
We hope that the presented analysis will inspire discussions and future work on characterizing heat and work in quantum processes.
Of course many open questions remain. One direction of particular relevance is clearly  the identification of a proper  \emph{metric} in configuration space,
that  would allow to quantify, in a precise and (hopefully) operationally well defined way, how distant two generic discrete trajectories are. 

\acknowledgements
VG thanks Rosario Fazio and Seth Lloyd for fruitful discussions and comments. 
VG is supported by MIUR through FIRB-IDEAS project No. RBID08B3FM.
JA is supported by the Royal Society.

\appendix

\section{Minimum and maximum heat for a single DUT+DTT process} \label{sec:min&max}

Here we discuss the impact of the unitary $V$ of Eq.~(\ref{eq:heatDUT+DTT}) on the heat of the DUT+DTT process~(\ref{eq:DUT+DTT}). Specifically, we want to identify the DUTs that maximize and minimize the heat on the r.h.s of Eq.~(\ref{eq:gg1113}) and establish if a DUT exists that leads to a saturated Clausius equality. The last question can be easily solved by noticing that saturation occurs only when the inequality holds in  Eq.~(\ref{eq:s-bound}). However, this is only true iff $\rho_1 = \rho_f$, i.e. $\rho_i$ and $\rho_f$ must be unitarily equivalent for some unitary $V_0$, $V_0 \rho_i V_0^{\dag} = \rho_f$. In other words equivalence in (\ref{eq:DUT+DTT}) requires that no DTT enters in the process so that  $\Delta S(\rho_i, \rho_f) = 0$. For any non-trivial DTT a finite gap between the l.h.s. and the r.h.s.
of Eq.~(\ref{eq:gg1113}) exists. (This  is not true however for sequences of DUT-DTT transformations as considered in Sec.~\ref{sec:sat}.)

\begin{figure}[t]
	\begin{center}
	{\includegraphics[width=\figsize\textwidth]{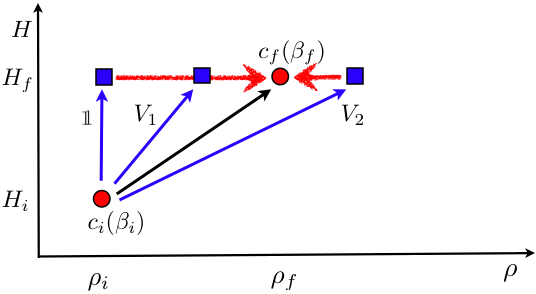}}
\caption{Realisations of the map from the Gibbs state  $c_i (\beta_i) = (\rho_i, H_i)_{\beta_i}$ to the Gibbs state $c_f (\beta_f) = (\rho_f, H_f)_{\beta_f}$ via some DUT, specified by a unitary $V$, followed by an appropriate DTT. }
\label{fig:max-minheat}
\end{center}
\end{figure}

To determine the minimum/maximum gap we require the maximum/minimum heat with respect to all possible DUTs of a DUT+DTT process connecting two thermal configurations $c_i (\beta_i) = (\rho_i, H_i)_{\beta_i}$ and $c_f (\beta_f) = (\rho_f, H_f)_{\beta_f}$, see Fig~\ref{fig:max-minheat}, i.e. 
\begin{equation}\label{eq:heatDUT+DTT1}
	\begin{split}
	Q_{\max} 
	&:=  \max_{V} \tr [(\rho_f - V \, \rho_i \, V^\dag) \; H_f ], \\
	Q_{\min} 
	&:=  \min_{V} \tr [(\rho_f - V \, \rho_i \, V^\dag) \; H_f ],
	\end{split}
\end{equation} 
where the maximisation/minimisation is taken over all unitary transformations $V$. This task is solved with the following Lemma.
\begin{lemma}
	Given $A=\sum_j \alpha_j |\alpha_j\rangle\langle \alpha_j|$ and $B=\sum_j \beta_j |\beta_j\rangle\langle \beta_j|$ Hermitian operators on a finite dimensional Hilbert space ${\cal H}$, with $\alpha_j$ and $\beta_j$ being their corresponding eigenvalues which are ordered in decreasing order (i.e. $\alpha_j \geqslant \alpha_{j+1}$, $\beta_j \geqslant \beta_{j+1}$). Then the minimum of $\tr [ A \, V \, B \, V^\dag]$ over the set of unitary transformations is achieved by the unitary $V_{\min}$ which maps the eigenvector $\{|\beta_j\rangle\} $ of $B$ into the eigenvectors $\{|\alpha_j\rangle\} $ of $A$ in such a way that 
\begin{eqnarray}
	V_{\min} |\beta_{j}\rangle = |\alpha_{N-j+1}\rangle ,
\end{eqnarray}  
i.e. the maximum eigenvector of $B$ is mapped into the minimum eigenvector of $A$. As a consequence the minimum expectation value is 
\begin{equation} \label{eq:min}
	\min_{V} \tr [ A \, V \,  B \, V^\dag] = \tr[ A \, V_{\min} \, B \, V_{\min}^\dag] = \sum_j \alpha_j \beta_{N-j+1}. 
\end{equation} 
Similarly the maximum  of $\tr [ A \, V \, B \, V^\dag]$ over the set of unitary transformations is achieved by the unitary $V_{\max}$  that maps the eigenvectors $\{|\beta_j\rangle\} $ of $B$ into the eigenvectors $\{|\alpha_j\rangle\} $ of $A$ in such a way that 
\begin{eqnarray}
	V_{\max} |\beta_{j}\rangle = |\alpha_{j}\rangle .
\end{eqnarray}  
Consequently the maximum expectation value is
\begin{equation}  \label{eq:max}
	\max_{V} \tr [ A \, V \, B \, V^\dag] 
	= \tr [ A \, V_{\max} \, B \, V_{\max}^\dag] = \sum_j \alpha_j \beta_{j}.
\end{equation} 
\end{lemma}
{\em Proof:} These minimum and maximum expectation values are a trivial consequence of the Theorem 2 of Ref.~\cite{LENARD}.   $\blacksquare$ \\ 

Application of Eqs.~(\ref{eq:min}) and (\ref{eq:max}) gives the minimum and maximum heat for the DUT+DTT process~(\ref{eq:DUT+DTT})
\begin{equation}\label{eq:heatDUT+DTT2}
	\begin{split}
	Q_{\max} 
	&=   \sum_{k=1}^N H_f(k) \Big(\frac{e^{- \beta_f H_f(k)}}{Z_f} -  \frac{e^{- \beta_i H_i(k)}}{Z_i} \Big),\\
   	Q_{\min} 
	&=   \sum_{k=1}^N  H_f(k) \Big(\frac{e^{- \beta_f H_f(k)}}{Z_f} -  \frac{e^{- \beta_i H_i(N-k+1)}}{Z_i}\Big),
	\end{split}
\end{equation} 
where $\{H_f(k)\}_k$ and $\{H_i(k)\}_k$ are the eigenvalues of $H_f$ and $H_i$ ordered in decreasing order.

\section{Proof of Theorem \ref{lem:Clausiussat}}  \label{sec:satproof}

Consider the $k$-th step of the trajectory ${\cal T}$ in panel {\bf b} in Fig.~\ref{fig:multiple}, which connects the thermal points  $c_{k} (\beta_k) = (\rho_{k}, H_{k})_{\beta_{k}}$ and $c_{k+1} (\beta_{k+1})=(\rho_{k+1},H_{k+1})_{\beta_{k+1}}$. We define a new trajectory ${\cal T}'_{k;n}$ which is identical with the original trajectory ${\cal T}$ except that the $k$-th step is now replaced with a sequence of $n-1$ intermediate \emph{thermal configurations} $c'_1(1)$,  $\cdots$,  $c'_{n-1} (1)$ that are linked through a DUQ-DTT sequence 
\begin{equation} 	\label{real22222}
	c_{k} \stackrel{DUQ+DTT}{\longrightarrow}  c'_1 
		\quad \cdots 
		\stackrel{DUQ+DTT}\longrightarrow c'_{n-1} 
		\stackrel{DUQ+DTT}\longrightarrow c_{k+1}.
\end{equation}
The configurations $c'_{\ell}(1)$ have density matrices $\rho'_1$, $\cdots$, $\rho'_{n-1}$  defined by the mixtures
\begin{equation} \label{rhoprime}
	{\rho}'_\ell = \left(1-\frac{\ell}{n}\right)\; {\rho}_{k} +\frac{\ell}{n}\; {\rho}_{k+1}
	\quad \mbox{for} \quad  \ell=0,\cdots, n,
\end{equation}
where $\rho_{0}'= \rho_{k}$ and $\rho'_{n}=\rho_{k+1}$. Applying the Clausius inequality~(\ref{yyy1}) in the form of (\ref{eq:SL}) to sequence~(\ref{real22222}) yields 
\begin{equation}  \label{yyy1121}
	\Delta S(\rho_{k}, \rho_{k+1}) 
						\geqslant \Lambda({\cal T}'_{k;n}).
\end{equation}
By definition (\ref{eq:Lambdapiece}), the transformations being \emph{discrete unitaries} and the fact that intermediate inverse temperatures are all set to 1, $\Lambda({\cal T}'_{k;n})$ can be expressed as
\begin{equation}
	\begin{split}
		 \Lambda({\cal T}'_{k;n}) 
		&=  \sum_{\ell=0}^{n-1}  \tr [(\rho'_{\ell} -{\rho}'_{\ell+1}) \; \ln \rho'_{\ell+1} ] \\
		&= \tr \left[ ( \rho_{{k}} - \rho_{k+1})\; \frac{1}{n}  \sum_{\ell=0}^{n-1} \;  \ln \rho'_{\ell+1} \right].
	\end{split}
\end{equation}
On the other hand the entropy change $\Delta S(\rho_{k}, \rho_{k+1})$ can be lower bounded by 
\begin{equation} \label{UPPERBOUND1}
 	\begin{split}
		\Delta S(\rho_{k}, \rho_{k+1}) 
		&=  \sum_{\ell=0}^{n-1}  \Delta S(\rho'_{\ell}, \rho'_{\ell+1}) \\
		& \leqslant       \sum_{\ell=0}^{n-1}  \tr [(\rho'_{\ell} -{\rho}'_{\ell+1}) \; \ln \rho'_{\ell} ] \\
 		&=  \tr \left[ ( \rho_{{k}} - \rho_{k+1})\; \frac{1}{n}  \sum_{\ell=0}^{n-1} \; \ln \rho'_{\ell} \right]\\
		&=  \Lambda({\cal T}_{k;n}')  + \tr \left[ ( \rho_{k} - \rho_{k+1})\;   \frac{\ln \rho_k - \ln \rho_{k+1}}{n} \right]\\
		&= \Lambda({\cal T}_{k;n}')  + \frac{S(\rho_{k+1}\| \rho_{k})+ S(\rho_{k}\| \rho_{k+1})}{n},
	\end{split}
\end{equation} 
which implies 
\begin{equation} \label{tttt}
	\Lambda({\cal T}_{k;n}') \geqslant  \Delta S(\rho_{k}, \rho_{k+1}) -  \frac{S(\rho_{k+1}\| \rho_{k})+ S(\rho_{k}\| \rho_{k+1})}{n}.
\end{equation} 
$\rho_k$ and $\rho_{k+1}$ are density matrices of Gibbs configurations and thus of full rank. Consequently, the quantity
 $S(\rho_{k+1}\| \rho_{{k}})+ S(\rho_{{k}}\| \rho_{k+1})$ is finite
\footnote{The symmetric sum of relative entropies $S(\rho_{k+1}\| \rho_{{k}})+ S(\rho_{{k}}\| \rho_{k+1})$ explodes if and only if  the kernel of $\rho_{k}$ or $\rho_{k+1}$ admits a non-trivial overlap with the support of the other state \cite{PETZ}. Since $\rho_k$ and $\rho_{k+1}$ are full rank, neither conditions can be fulfilled and the symmetric sum of relative entropies is always finite.}.
From Eq.~(\ref{yyy1121}) and (\ref{tttt}) it then follows that $\Lambda({\cal T}_{k;n}')$ converges to $\Delta S(\rho_{k}, \rho_{k+1})$ for $n \to \infty$, i.e. 
\begin{eqnarray}  \label{limit}
	\Delta S(\rho_{k}, \rho_{k+1}) = \lim_{n\rightarrow \infty} 	\Lambda({\cal T}_{k;n}').
\end{eqnarray}
In other words by augmenting the intermediate points of which connects $c_{k} (\beta_k)$ and $c_{k+1} (\beta_{k+1})$ we can saturate the associated Clausius inequality for the $k$-th step of the trajectory ${\cal T}$. By repeating the same procedure for each of  the steps  of ${\cal T}$ a new trajectory emerges as the union of the individual sequences, 
\begin{equation}
	{\cal T}'_{n} = \bigcup_{k=1}^{N-1}   {\cal T}_{k;n_k}'
\end{equation}
where $n$ is the multidimensional variable $(n_1,n_2, \cdots,  n_{N-1})$. Lemma \ref{lem:Clausiussat} follows from the additivity of $\Lambda$ (\ref{functionalS}) and taking the limit of each $n_k \to \infty$ ~(\ref{limit}). $\blacksquare$


\end{document}